\documentclass[aps,pra,superscriptaddress,floatfix,showpacs,10pt, twocolumn,hidekeys]{revtex4}
\usepackage{bbm}
\usepackage{indentfirst}
\usepackage{amsfonts}
\usepackage[dvips]{graphicx}
\usepackage{amsmath,amsfonts,amssymb,graphics,graphicx,epsfig,color,times,bbm}

\usepackage{lipsum}
\usepackage{graphicx}
\usepackage{epstopdf}
\usepackage{subfigure}
\if CLASSOPTIONcompsoc
\usepackage[caption=false, font=normalsize, labelfont=sf, textfont=sf]{subfig}
\else
\usepackage[caption=false, font=footnotesize]{subfig}
\fi
\DeclareMathOperator{\Tr}{Tr}

\begin{document}

\title{Entanglement Dynamics in Anti-$\mathcal{PT}$-Symmetric Systems}
\author{Yu-Liang Fang}
\thanks{Yu-Liang Fang and Jun-Long Zhao contributed equally to this work}
\affiliation{Quantum Information Research Center, Shangrao Normal University, Shangrao 334001, China}

\author{Jun-Long Zhao}
\thanks{Yu-Liang Fang and Jun-Long Zhao contributed equally to this work}
\affiliation{Quantum Information Research Center, Shangrao Normal University, Shangrao 334001, China}

\author{Dong-Xu Chen}
\affiliation{Quantum Information Research Center, Shangrao Normal University, Shangrao 334001, China}

\author{Yan-Hui Zhou}
\affiliation{Quantum Information Research Center, Shangrao Normal University, Shangrao 334001, China}

\author{Yu Zhang}
\thanks{smutnauq@smail.nju.edu.cn}
\affiliation{Quantum Information Research Center, Shangrao Normal University, Shangrao 334001, China}
\affiliation{School of Physics, Nanjing University, Nanjing 210093, China}

\author{Qi-Cheng Wu}
\thanks{wqc@sru.edu.cn}
\affiliation{Quantum Information Research Center, Shangrao Normal University, Shangrao 334001, China}

\author{Chui-Ping Yang}
\thanks{yangcp@hznu.edu.cn}
\affiliation{Quantum Information Research Center, Shangrao Normal University, Shangrao 334001, China}
\affiliation{School of Physics, Hangzhou Normal University, Hangzhou 311121, China}

\author{Franco Nori}
\thanks{fnori@riken.jp}
\affiliation{Theoretical Quantum Physics Laboratory, RIKEN, Wako-shi, Saitama 351-0198, Japan}
\affiliation{RIKEN Center for Quantum Computing (RQC), Wako-shi, Saitama 351-0198, Japan}
\affiliation{Physics Department, The University of Michigan, Ann Arbor, Michigan 48109-1040, USA}


\begin{abstract}
  In the past years, many efforts have been made to study various noteworthy phenomena in both parity-time ($\mathcal{PT}$) and anti-parity-time ($\mathcal{APT}$) symmetric systems. However, entanglement dynamics in $\mathcal{APT}$-symmetric systems has not previously been investigated in both theory and experiments. Here, we investigate the entanglement evolution of two qubits in an $\mathcal{APT}$-symmetric system. In the $\mathcal{APT}$-symmetric unbroken regime, our theoretical simulations demonstrate the periodic oscillations of entanglement when each qubit evolves identically, while the nonperiodic oscillations of entanglement when each qubit evolves differently. In particular, when each qubit evolves near the exceptional point in the $\mathcal{APT}$-symmetric unbroken regime, there exist entanglement sudden vanishing and revival. Moreover, our simulations demonstrate rapid decay and delayed death of entanglement provided one qubit evolves in the $\mathcal{APT}$-symmetric broken regime.  In this work, we also perform an experiment with a linear optical setup. The experimental results agree well with our theoretical simulation results. Our findings reveal novel phenomena of entanglement evolution in the $\mathcal{APT}$-symmetric system and opens a new direction for future studies on the dynamics of quantum entanglement in multiqubit $\mathcal{APT}$-symmetric systems or other non-Hermitian quantum systems.
\end{abstract}

\maketitle

\section{Introduction}
Since the discovery of the parity-time ($\mathcal{PT}$)-symmetric Hamiltonian in 1998 \cite {Bender},  this new class of non-Hermitian Hamiltonians has attracted extensive attention. It was found \cite{Konotop,Ganainy,gong2018,PengB,JingH} that $\mathcal{PT}$-symmetric Hamiltonians allow for real eigenvalues associated with observable quantities in numerous physical systems. {However, the intended real eigenvalues may fade, since there exist exceptional points \cite{ZhangGQ} where the $\mathcal{PT}$ symmetry is spontaneously broken and the eigenvalues become complex.} This striking feature has inspired numerous theoretical and experimental studies on non-Hermitian systems; and $\mathcal{PT}$ symmetry has been realized in both classical \cite{ZhangJ,Ozdemir} and quantum systems \cite{LiJ,TangJS,WangYT,XiaoL2,ArkhipovII}. Meanwhile, many remarkable quantum phenomena in $\mathcal{PT}$-symmetric systems have been explored, such as critical phenomena \cite{yashida8}, increase of entanglement \cite{slchen90}, chiral population transfer \cite{hxdason537,jdopp537}, decoherence dynamics \cite{bgardas94}, and information retrieval and criticality \cite{Kawabata2}. Theoretically, in $\mathcal{PT}$-symmetric systems, Ref. \cite{scaks063846} investigated {the} delayed sudden vanishing of entanglement at exceptional points,  \cite{BartkowiakM,WangX} studied entanglement sudden vanishing, and \cite{fmwn} realized effective entanglement recovery via operators. Entanglement, precision metrology, and sensing enhancement were reported in $\mathcal{PT}$-symmetric systems {\cite{ctpdsg,msdcl,LiuZP,CrokeS,BianZ,KalozoumisPA,PaivaD}}. Recent experiments have demonstrated topological edge states based on entanglement in $\mathcal{PT}$-symmetric quantum walks \cite{lxiaoxzhan}, stable states with nonzero entropy under broken $\mathcal{PT}$ symmetry \cite{jwwen3}, and optomechanical dynamics under the $\mathcal{PT}$- and broken-$\mathcal{PT}$-symmetric regimes \cite{XuH}.

\indent On the other hand, another important counterpart, anti-parity-time ($\mathcal{APT}$) symmetry, has recently attracted considerable interest. {Generally speaking, the Hamiltonian $\hat{H}_\mathcal{PT}$ in a $\mathcal{PT}$-symmetric system has a counterpart $\hat{H}_\mathcal{APT}$ in an $\mathcal{APT}$-symmetric system, and they are related by $\hat{H}_{\mathcal{APT}}=i\hat{H}_{\mathcal{PT}}$, i.e., there is a one-to-one correspondence between the Hamiltonians for $\mathcal{PT}$ and $\mathcal{APT}$ systems. However, although these two Hamiltonians differ only by an imaginary number, the dynamic properties of $\mathcal{APT}$-symmetric systems might be quite different from those of $\mathcal{PT}$-symmetric systems. Recently, many research groups \cite{YangY,LiY,hlzhang7594,fxzhang053901,QinY} have focused on the study of the dynamic properties of an $\mathcal{APT}$-symmetric system. For example, Choi \emph{et al.} \cite{Choi} showed that $\mathcal{PT}$-symmetric systems reveal distinct symmetry under a $\mathcal{PT}$ operation, while such symmetry does not exist in $\mathcal{APT}$-symmetric systems under the same $\mathcal{PT}$ operation. Moreover, the physical mechanisms of energy-exchange between environment and system are different for $\mathcal{APT}$- and $\mathcal{PT}$-symmetric cases. In other words, the $\mathcal{APT}$- and $\mathcal{PT}$-symmetric systems reveal different dynamic behavior (see Appendix A). Therefore, a comprehensive study of the dynamic properties of $\mathcal{APT}$-symmetric systems is important and interesting for fully understanding the features of open quantum systems \cite{Jahromi,GeL,YangF,Konotop2,KeS,LiQ}.}

Recently, it was found \cite{hlzhang7594,fxzhang053901} that exceptional points also exist in $\mathcal{APT}$-symmetric systems, where a phase transition occurs when the eigenvalues change from real ($\mathcal{APT}$-symmetric unbroken phase) to imaginary ($\mathcal{APT}$-symmetric broken phase). Moreover, witness of non-Markovianity can indicate the presence of exceptional points and behavior in $\mathcal{APT}$-symmetric systems, especially in high-dimensional qudits \cite{Jahromi}. In addition, several interesting phenomena have been reported in $\mathcal{APT}$-symmetric systems, such as optical materials with balanced positive and negative index \cite{GeL} and optical systems with constant refraction \cite{YangF}. Experimentally, $\mathcal{APT}$-symmetric systems have been realized in optics {\cite{KeS,Konotop2,LiQ,jzyllwck,ZhengC}}, atoms {\cite{ChuangYL,yjymyz}, } electrical circuit resonators \cite{Choi}, diffusive systems \cite{LiY}, and waveguides \cite{QinY}. Moreover, Ref. \cite{hlzhang7594} studied $\mathcal{APT}$-symmetry and observed its spontaneous breaking in a linear device by spinning a lossy resonator.  Reference \cite{PengP} demonstrated the first experiment of optical $\mathcal{APT}$ symmetry in a warm atomic-vapor cell. { Reference \cite{CenJ} showed that non-Hermitian $\mathcal{PT}$- and $\mathcal{APT}$-symmetric systems may be more suitable than conventional Hermitian systems for quantum computing and quantum information processing.} Experiments \cite{ZhangXL,GaoT} showed dynamically encircling of an exceptional point. Reference \cite{FangYL} experimentally demonstration of coherence flow in $\mathcal{PT}$- and $\mathcal{APT}$- symmetric systems.

\indent Quantum entanglement lies at the heart of quantum mechanics and is a core resource for potential applications in quantum information, quantum communication, and computation. The longevity of the available entanglement is of importance. It is well understood that any unavoidable interaction between a quantum system and an external environment brings noise to the system, which is substantially detrimental to the entanglement in the system. Particularly, Ref. \cite{tyu264} pointed out that, when a quantum system interacts with its surroundings, quantum entanglement created in the system will decay, and entanglement sudden vanishing may occur within a finite time, under the influence of environmental noise. This vanishing of entanglement has been both theoretically predicted and experimentally verified in a wide range of physical systems,  such as optical systems \cite{jsxu104,asc326}, effective solid-state spin baths in a diamond sample \cite{fwang98}, and atomic ensembles \cite{jlaurat99}. However, to the best of our knowledge, the entanglement dynamics in $\mathcal{APT}$-symmetric systems  has \emph{not} been theoretically and experimentally investigated. The study of entanglement evolution in $\mathcal{APT}$-symmetric systems is significant, which can uncover various phenomena that are different from Hermitian quantum systems, and reveal the relationship between non-Hermitian systems and their environments.

\indent In this work, we investigate the entanglement evolution of two qubits in an $\mathcal{APT}$-symmetric system. Noteworthy phenomena are found through theoretical simulations. First, when each qubit evolves identically in the $\mathcal{APT}$-symmetric unbroken regime, there exist periodic oscillations of the entanglement; while there exist nonperiodic oscillations of entanglement when each qubit evolves differently in the  $\mathcal{APT}$-symmetric unbroken regime. Second, when each qubit evolves near the exceptional point in the $\mathcal{APT}$-symmetric unbroken regime, there exist entanglement sudden vanishing and revival. Finally, the entanglement undergoes a rapid decay and a delayed vanishing provided one qubit evolves in the $\mathcal{APT}$-symmetric broken regime. We also perform an experiment with a linear optical setup. The experimental results agree well with our theoretical simulation results.

\section{Theory and experimental setup}
The non-Hermitian Hamiltonian of two qubits ($1$, $2$) in an $\mathcal{APT}$-symmetric system is given by \cite{LiY,Wen} (assuming $\hbar=1$)
\begin{equation}
\hat{H}=\hat{H}_{1,\mathcal{APT}}+\hat{H}_{2,\mathcal{APT}}.\label{Hapt2}
\end{equation}
Here, {$\hat{H}_{j,\mathcal{APT}}=i\gamma_j \hat{\sigma}_{j,x}+s_j\hat{\sigma}_{j,z}=\gamma_j \left(i \hat{\sigma}_{j,x}+a_j \hat{\sigma}_{j,z}\right)$} is the $\mathcal{APT}$-symmetric Hamiltonian of qubit $j$ $(j=1,2)$, and $\hat{\sigma}_{j,x}$ and $\hat{\sigma}_{j,z}$ are the standard Pauli operators. {Generally, $s_j\hat{\sigma}_{j,z}$ and $i \gamma_j \hat{\sigma}_{j,x}$ are the Hermitian and non-Hermitian parts of the Hamiltonian, respectively \cite{JuCY}; $\gamma_j$ is an energy scale, and $a_j=s_j/\gamma_j>0$ represents the degree of Hermiticity.} The eigenvalues of $\hat{H}_{j,\mathcal{APT}}$ are
\begin{equation}
 \lambda_{j, \mathcal{APT}} = \pm \gamma_j \sqrt{a_j^2-1},
\end{equation}
which are imaginary numbers when $0<a_{j}<1$ (the $\mathcal{APT}$-symmetric broken regime) \cite{XiaoL2,Wen}, and real numbers when $a_j>1$ (the $\mathcal{APT}$-symmetric unbroken regime). {Note that the eigenvalues of the Hamiltonian $\hat{H}_{j,\mathcal{APT}}$ are zero when $a_{j}=1$ (the exceptional point). For simplicity, we set $\gamma_j=1~(j=1,~2)$. Thus, the $\mathcal{APT}$-symmetric Hamiltonian of the $j$th qubit can be written as
\begin{equation}
\hat{H}_{j,\mathcal{APT}}=i \hat{\sigma}_{j,x}+a_j\hat{\sigma}_{j,z}.
\end{equation}}
  For the Hamiltonian $\hat{H}$, the time-evolution nonunitary operator $\boldsymbol{U}(t)=\rm{exp}(-i \hat{\emph{H}} \emph{t})$ is expressed as
 \begin{equation}
\boldsymbol{U}(t)=\boldsymbol{U}_{1,\mathcal{APT}}(t)\otimes\boldsymbol{U}_{2,\mathcal{APT}}(t),\label{UAPT}
 \end{equation}
where
  \begin{equation}
  \boldsymbol{U}_{j,\mathcal{APT}}(t)=\exp \left(-i\hat{H}_{j,\mathcal{APT}}t\right) , ~~j=1,~2.\label{UAPT12}
  \end{equation}
 Equation (\ref{UAPT12}) is the time-evolution nonunitary operator of qubit $j$ in the $\mathcal{APT}$-symmetric system. In our experiment, we realize the operator $\boldsymbol{U}(t)$  by a combination of optical elements, and access the time-evolved states by enforcing $\boldsymbol{U}(t)$ on the initial states. The optical simulation of the time-evolution nonunitary operator $\boldsymbol{U}(t)$ is shown in the gray part in Fig.~\ref{fig1}. The operator $\boldsymbol{U}_{j,\mathcal{APT}}$ can be expressed as (see Appendix B):
 \begin{equation}
 \boldsymbol{U}_{j,\mathcal{APT}}  = U_{j,2}(\theta_{j,2})L_{j}(\xi_{j,1},\xi_{j,2})U_{j,1}(\theta_{j,1}),\label{Eq:UaptE}
 \end{equation}
 with
  \begin{align*}
  U_{j,1}(\theta_{j,1})=&R_{\rm HWP}(0^{\circ})R_{HWP}(22.5^{\circ})R_{\rm QWP}(45^{\circ})\\
  &R_{\rm HWP}(\theta_{j,1}) R_{\rm QWP}(45^{\circ}),\\
  U_{j,2}(\theta_{j,2})=&R_{\rm QWP}(45^{\circ}) R_{\rm HWP}(\theta_{j,2}) R_{\rm QWP}(45^{\circ}) \\
  &R_{\rm HWP}(67.5^{\circ}),\\
  L_{j}(\xi_{j,1},\xi_{j,2})=&\left(\begin{array}{cc}
0 & \sin 2 \xi_{j,1} \\
\sin 2 \xi_{j,2} & 0
\end{array}\right),
  \end{align*}
  \begin{figure*}[!tbp]
  \setlength{\belowcaptionskip}{-0.6cm}
  \begin{flushleft}\hspace*{0pt}
  \centerline{
  \includegraphics[width=1\linewidth]{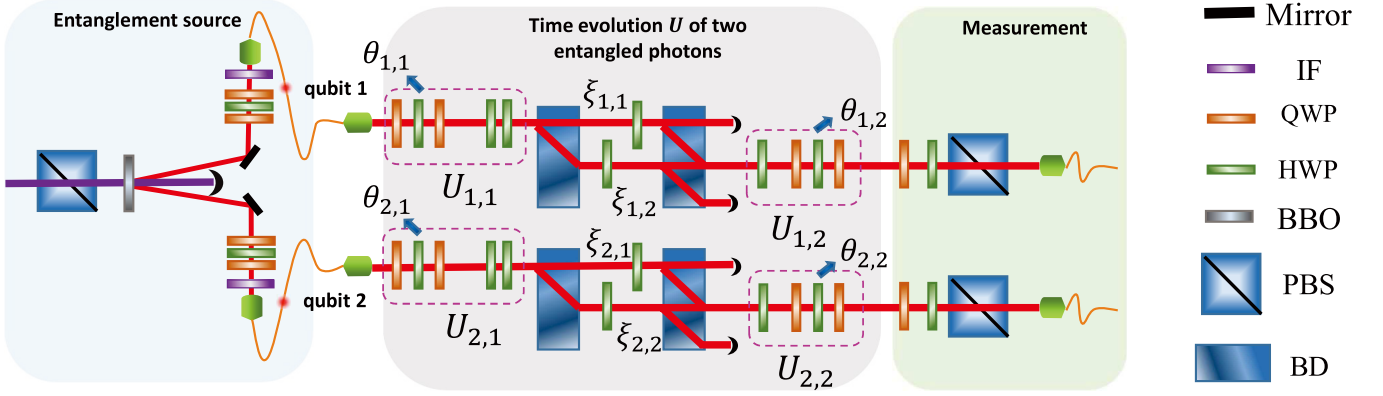}}
  \caption{Experimental setup. Two entangled 808 nm photons are generated through degenerate spontaneous parametric down-conversion, by pumping two type-II phase matched nonlinear $\beta$-barium-borate (BBO) crystals with the 404 nm pump laser shown in the {far left}. In the time evolution part (center area), the optical elements in the upper (lower) layer are used to realize the nonunitary time evolution operator $U_{\mathcal{APT},1}$ $( U_{\mathcal{APT},2})$. The wave-plates in the four dashed rectangles are used to implement the operators $U_{1,1}$, $U_{1,2}$, $U_{2,1}$, and $U_{2,2}$, respectively; while the upper (lower) two beam displacers (BDs) with two HWPs inside are used to realize the loss operator $L_1$ ($L_2$). In the measurement part (right area), the density matrix is reconstructed through quantum state tomography.}
 \label{fig1}
  \end{flushleft}
 \end{figure*}
where $R_{\rm HWP}\left(R_{\rm QWP}\right)$ is the rotation-operation operator of a half-wave plate (quarter-wave plate) and $L_{j}(\xi_{j,1},\xi_{j,2})$ is the loss-dependent operator. Thus, the nonunitary dynamics of the $\mathcal{APT}$-symmetric system is captured by the time-dependent density matrix \cite {Kawabata2,XiaoL2,brody109}
\begin{equation}\label{pure}
\boldsymbol{\rho}(t)=\frac{\boldsymbol{U}(t) \boldsymbol{\rho}(0) \boldsymbol{U}^{\dagger}(t)}{\Tr\left[\boldsymbol{U}(t) \boldsymbol{\rho}(0) \boldsymbol{U}^{\dagger}(t)\right]}.
\end{equation}
\indent In our experiment, the two qubits are two photons each having two orthogonal polarized states $|H\rangle$ and $|V\rangle$. As shown in Fig.~\ref{fig1}, the initial two-photon entangled Bell state $|\phi_{0}\rangle=(|HV\rangle+|VH\rangle)/\sqrt{2}$ is generated via a spontaneous parametric down-conversion process (left area), then each photon experiences an independent time evolution. Here, the sandwich structure device (QWP-HWP-QWP) is introduced to compensate the phase between the photons. Experimentally, we reconstruct the density matrix at any given time $t$ via quantum state tomography after the two photons passing through the time evolution section. Essentially, we project the photons onto 16 bases through a combination of QWP, HWP, and PBS, and then perform a maximum-likelihood estimation of the density matrix (Tomography) \cite{DFVJ64}. The measurement of the photon source  yields a maximum of 10,000 photon counts over 1.5 s after the 10 nm interference filter (IF).\\
\indent We here adopt the concurrence as a measure of entanglement. The concurrence is calculated as \cite{wkw2245}:
\begin{equation}
C(\boldsymbol{\rho})=\max \left\{0, \sqrt{\lambda_{1}}-\sqrt{\lambda_{2}}-\sqrt{\lambda_{3}}-\sqrt{\lambda_{4}}\right\},\label{concurrence}
\end{equation}
where $\lambda_{j}$ $(j=1,~2,~3,~4)$ are the eigenvalues of the matrix $R=\rho\left(\sigma_{y} \otimes \sigma_{y}\right) \rho^{*}\left(\sigma_{y} \otimes \sigma_{y}\right)$ in decreasing order, and $\sigma_{y}$ is the Pauli $y$ matrix. In this work, the theoretical and experimental concurrences are both calculated using Eq.~(\ref{concurrence}). For the experimental concurrence, the density matrix $\rho$ is reconstructed by quantum state tomography. While, for the theoretical concurrence, $\rho$ is determined by numerically solving Eq.~(\ref{pure}) based on the operator $\boldsymbol{U}(t)$ in Eq.~(\ref{UAPT}) and the Hamiltonian $\hat{H}$ in Eq.~(\ref{Hapt2}). For the initial two-photon Bell state $|\phi_0\rangle$, the concurrence is maximal, i.e., $C_{\max}(\boldsymbol{\rho})=1$.

\section{Theoretical simulations and experimental results}
\begin{figure*}[!htbp]
\setlength{\belowcaptionskip}{0cm}
    \subfigure{}
       \includegraphics[width=0.46\linewidth]{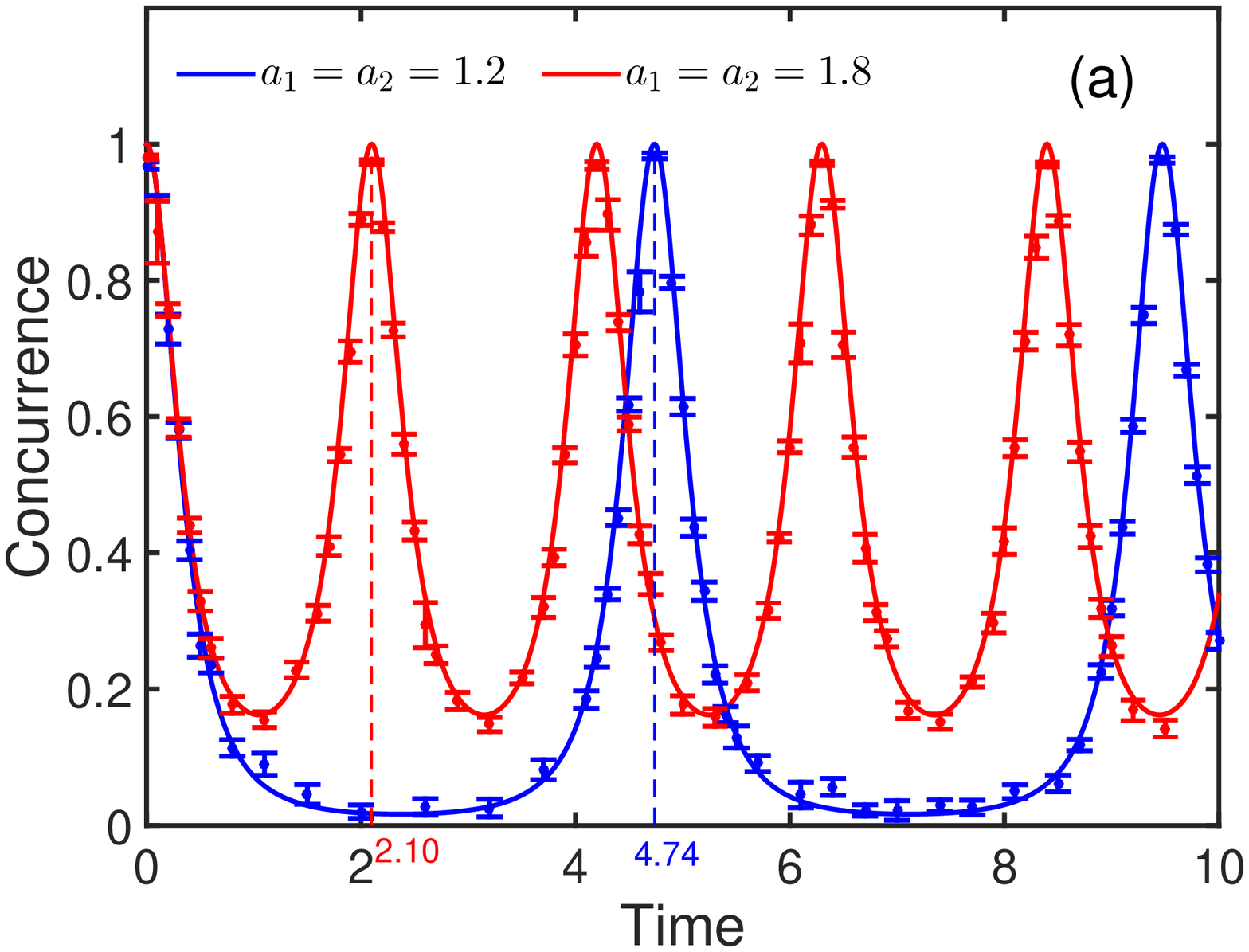}\hspace*{15pt}
    \subfigure{}
        \includegraphics[width=0.46\linewidth]{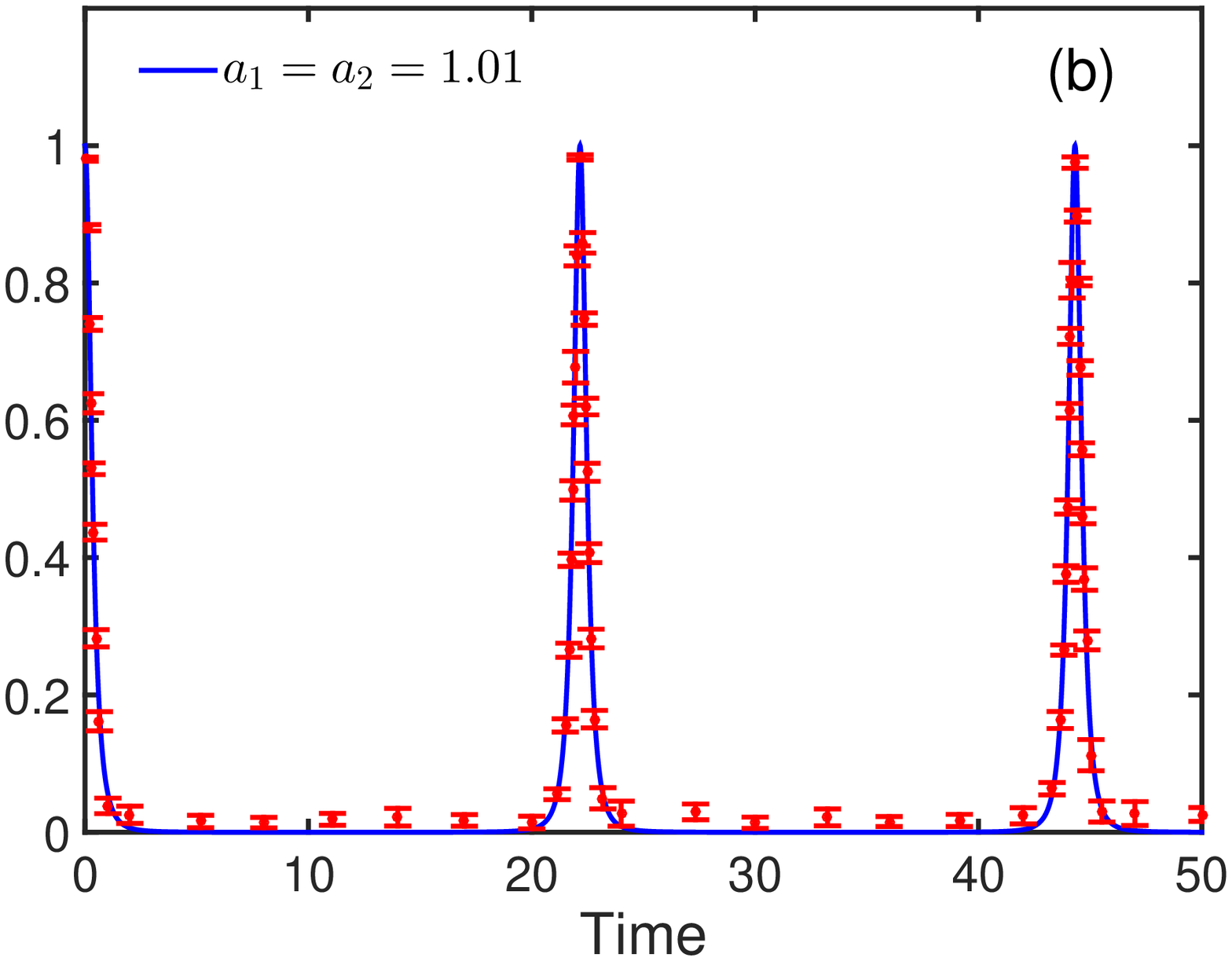}
\caption{Time evolution of the concurrence when the two qubits evolve identically in the $\mathcal{APT}$-symmetric unbroken regime. (a) $a_{1}=a_{2}=1.2$ (blue curve), $a_{1}=a_{2}=1.8$ (red curve). (b) $a_{1}=a_{2}= 1.01$ (close to the exceptional point). All curves are theoretical results while the dots are experimental data. {Note that the unit of time here is $1/\gamma_j$, which also applies to Figs.~(\ref{fig3},~\ref{fig4},~\ref{afig4},~\ref{afig5}).}}\label{fig2}
\end{figure*}
We first consider the case when both qubits evolve identically, i.e. $a_{1}=a_{2}$. In Fig. \ref{fig2}, we investigate the time-evolution dynamics of entanglement when the two qubits evolve in the $\mathcal{APT}$-symmetric unbroken regime $(a_{1}=a_{2}>1)$. In Fig. \ref{fig2}(a), we plot the evolution of entanglement when: (i) $a_{1}=a_{2}=1.2$ (blue curve), (ii) $a_{1}=a_{2}=1.8$ (red curve). One can see that the concurrence oscillates periodically over time $t$, and the minimum values of the concurrence for the two cases are 0.016 and 0.162, respectively; while the peak value for the two cases is 1. Furthermore, the oscillation period increases with decreasing the parameter $a_{1}(a_{2})$. In Fig.~\ref{fig2}(b), we further investigate the phenomena near the exceptional point $(a_{1}=a_{2}= 1.01)$. {It is clear that the concurrence still changes periodically over time, but drops close to zero and then rises after a non-zero time duration. Thus, to some extent, the sudden vanishing and revival of concurrence occurs.} In addition, combining Fig.~\ref{fig2}(a) with Fig.~\ref{fig2}(b), we can confirm that the concurrence changes periodically over time and the oscillation period increases when decreasing the parameter $a_{1}(a_{2})$. {Thus, in order to clearly show the dynamics of the concurrence in the vicinity of the exceptional point, we have chosen a larger time range for Fig. \ref{fig2}(b). }

\begin{figure*}[!htbp]
\setlength{\belowcaptionskip}{-0cm}
    \subfigure{}
       \includegraphics[width=0.45\linewidth]{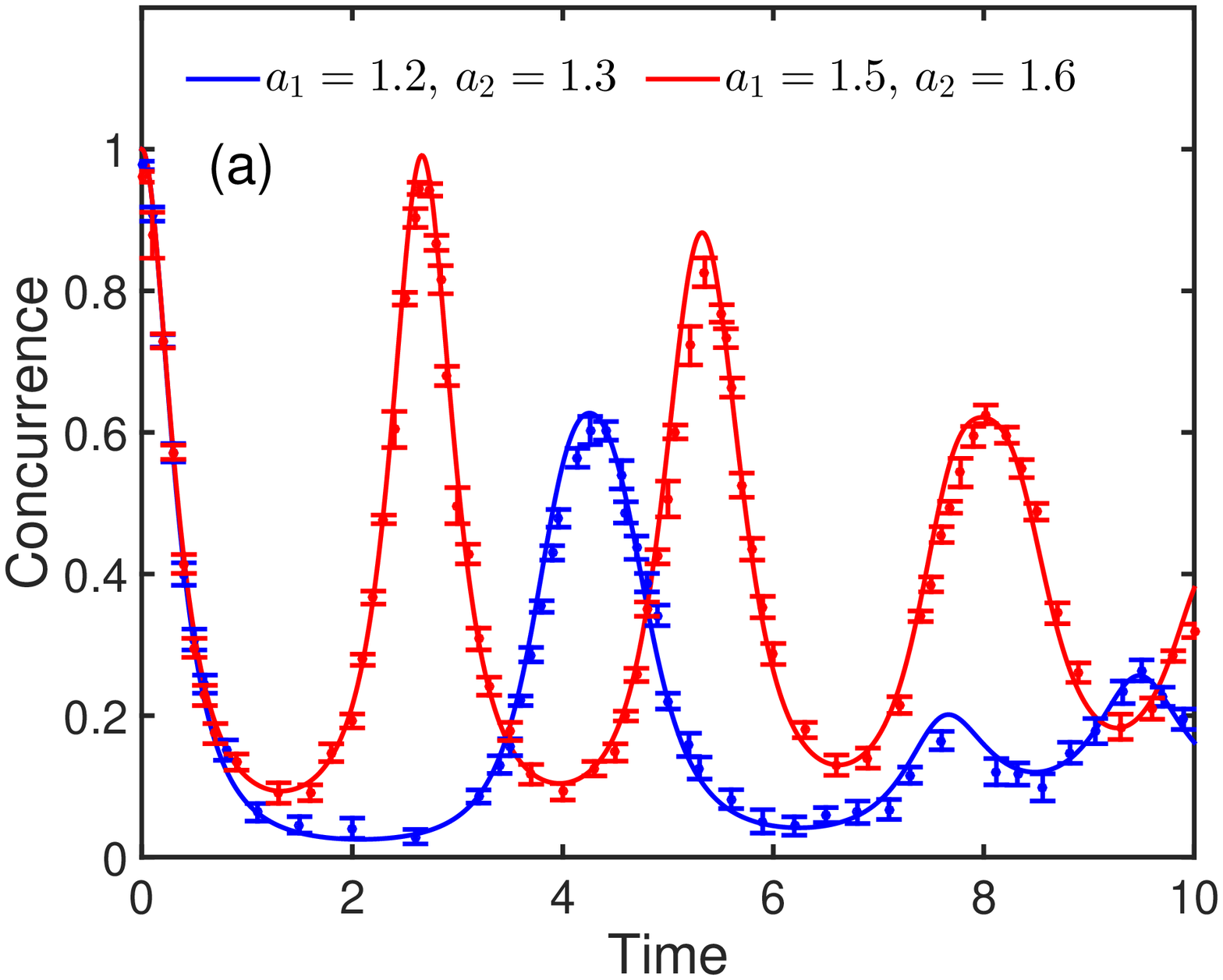}\hspace*{15pt}
    \subfigure{}
        \includegraphics[width=0.43\linewidth]{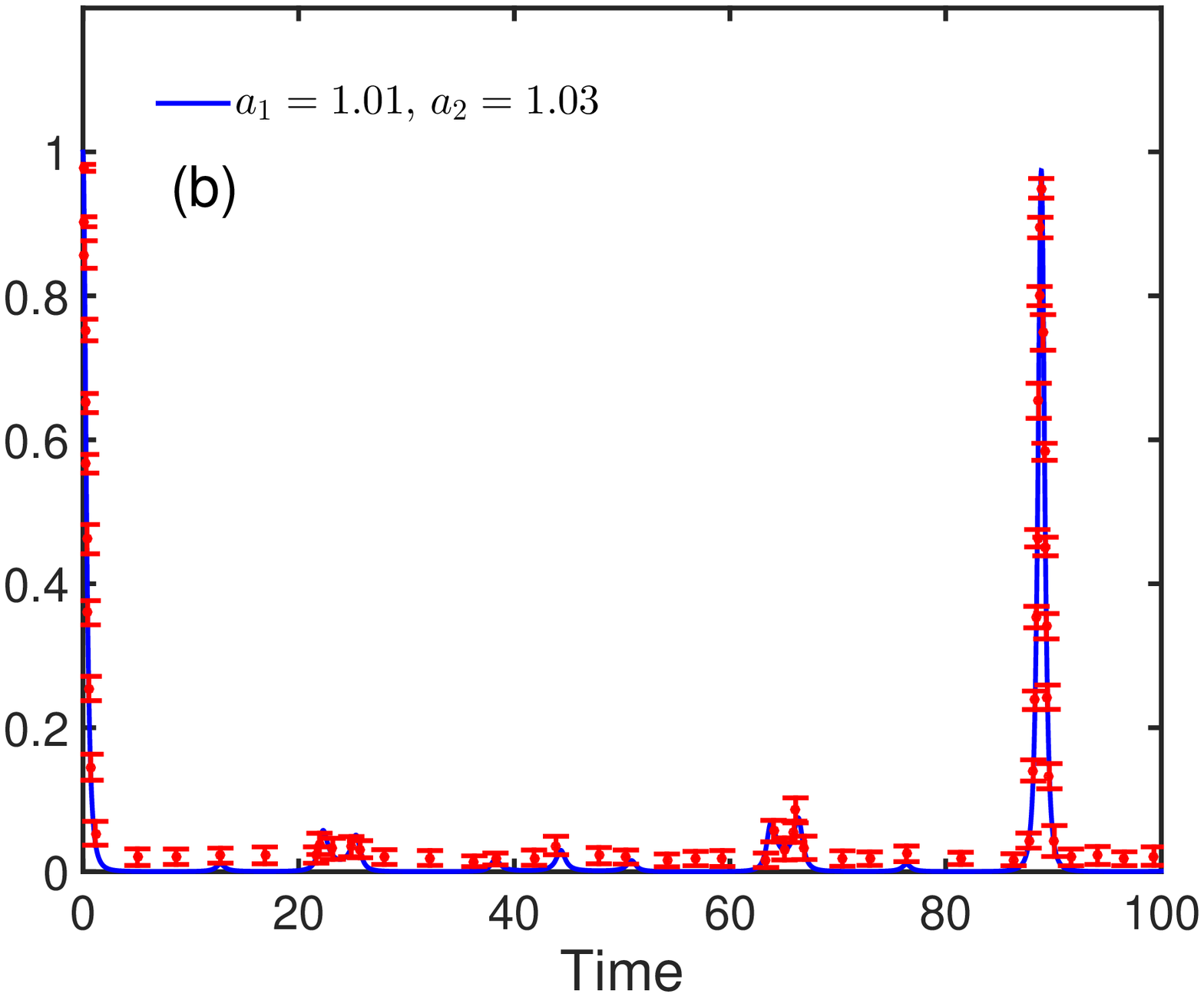}
\caption{ Time evolution of the concurrence when the two qubits evolves differently in the $\mathcal{APT}$-symmetric unbroken regime. (a) $a_{1}=1.2, a_{2}=1.3$ (blue curve), $a_{1}=1.5, a_{2}=1.6$ (red curve). (b) $a_1=1.01, a_2=1.03$ (close to the exceptional point). All curves are theoretical results while the dots are experimental data.}\label{fig3}
\end{figure*}

Next, we further investigate the relationship between the period of concurrence and the  parameter  $a = a_1 = a_2$. As shown in  Appendix C,  we find that the oscillation period is $\frac{\pi}{\sqrt{a^2-1}}$, where $a  > 1$. Here, we also provide some physical explanations for the above theoretical relationship and experimental  results. Since $a_1$($a_2$) represents the degree of Hermiticity of the Hamiltonian,  the ratio of the Hermitian component $a_1\sigma_{1,z}$ ($a_2\sigma_{2,z}$) to the non-Hermitian component $i\sigma_{1,x}$ ( $i\sigma_{2,x}$) in the Hamiltonian $H_{1,\mathcal{APT}}$ ($H_{2,\mathcal{APT}}$) increases when increasing $a_1$ ($a_2$).  In the limiting case $a_1 = a_2\gg 1$, the $\mathcal{APT}$ -symmetric system reverts to a Hermitian system, and thus the entanglement inherent between the two qubits remains unchanged. On the other hand, increasing the parameter $a_1$ ($a_2$) also implies increasing the Hermitian component $a_1\sigma_{1,z}$ ($a_2\sigma_{2,z}$). In this case, the energy of the $\mathcal{APT}$-symmetric system increases accordingly, which leads to a faster time evolution behavior with smaller oscillation period.

\indent In the above, we have considered the case when both qubits evolve identically. Next we consider a more general case when the two qubits evolve in different ways,  i.e., $a_1 \neq a_2$. In Fig.~\ref{fig3}, we investigate the time-evolution dynamics of entanglement when the two qubits evolve in the $\mathcal{APT}$-symmetric unbroken regime ($a_1 \neq a_2>1$). In Fig.~\ref{fig3}(a), we plot the evolution of entanglement for: (i) $a_1=1.2$, $a_2=1.3$ (blue curve); (ii) $a_1=1.5$, $a_2=1.6$ (red curve). From Fig.~\ref{fig3}(a), one can see that there exists nonperiodic oscillations of entanglement when $a_1 \neq a_2$. {The  parameter $a_1$ ($a_2$) represents  the degree of Hermiticity of the Hamiltonian. When decreasing the parameter $a_1$ ($a_2$), the non-Hermitian component $i\sigma_{1,x}$ ($i\sigma_{2,x}$)  gradually plays a dominant role in the time evolution.  In this case, the dissipative coupling between system and environment increases,  and the quantum information carried by the system gradually flows to the environment. Thus,  the concurrence of  the  $\mathcal{APT}$ -symmetric system will gradually  decrease.} In Fig.~\ref{fig3}(b), we investigate these phenomena near the exceptional point ($a_1=1.01$, $a_2=1.03$). Figure \ref{fig3}(b) shows that the concurrence varies over time nonperiodically, accompanied with the sudden vanishing and revival of entanglement. Nonperiodic oscillations in Fig.~\ref{fig3} show the entanglement loss and recovery, which is observed experimentally and agrees with the theoretical results.

\begin{figure*}[!htbp]
\setlength{\belowcaptionskip}{-0cm}
   \begin{flushleft}\hspace*{-0.4pt}
		\subfigure{}
		\includegraphics[width=0.46\linewidth]{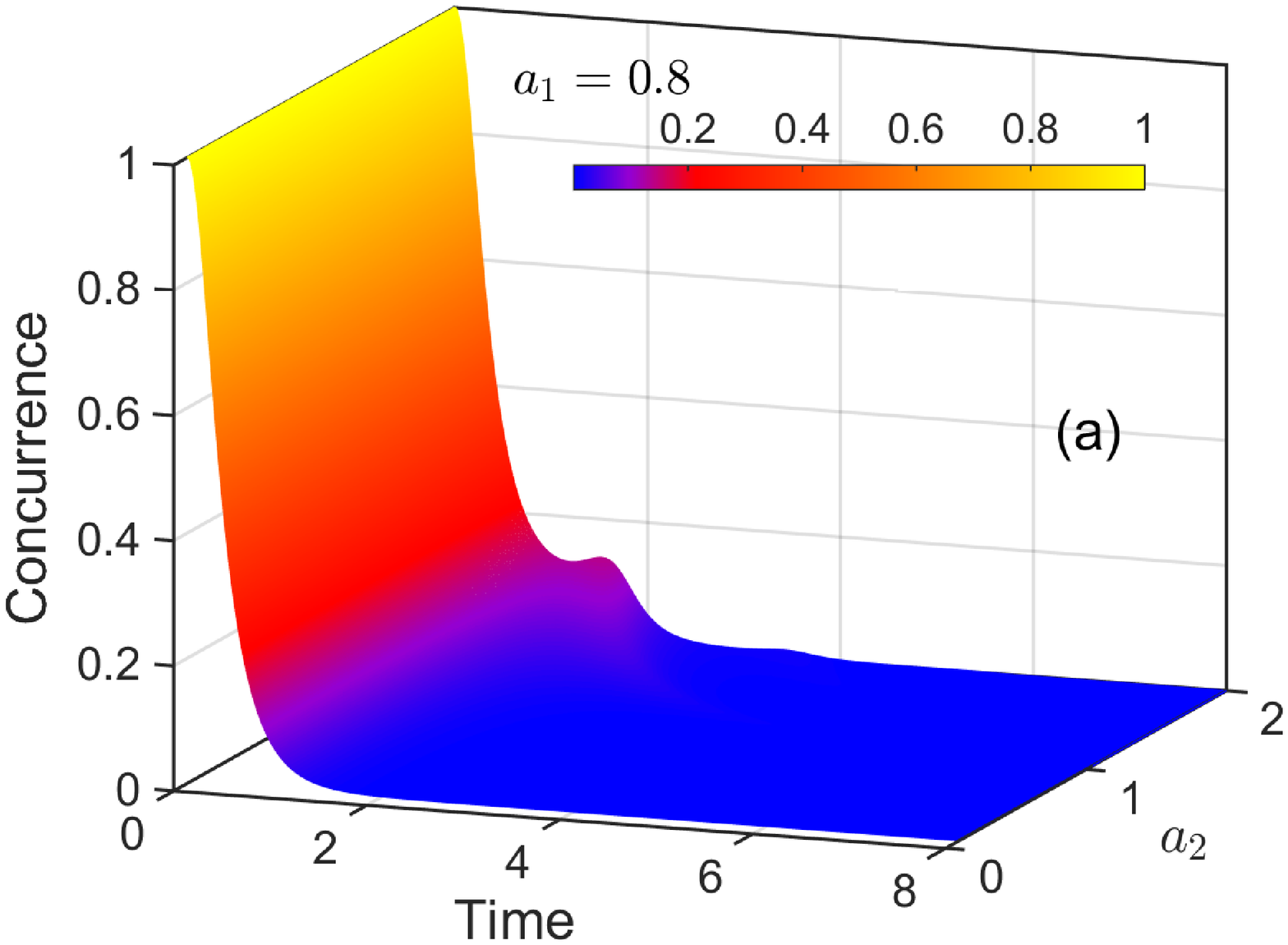}\hspace*{15pt}
		~~
		\subfigure{}
		\includegraphics[width=0.46\linewidth]{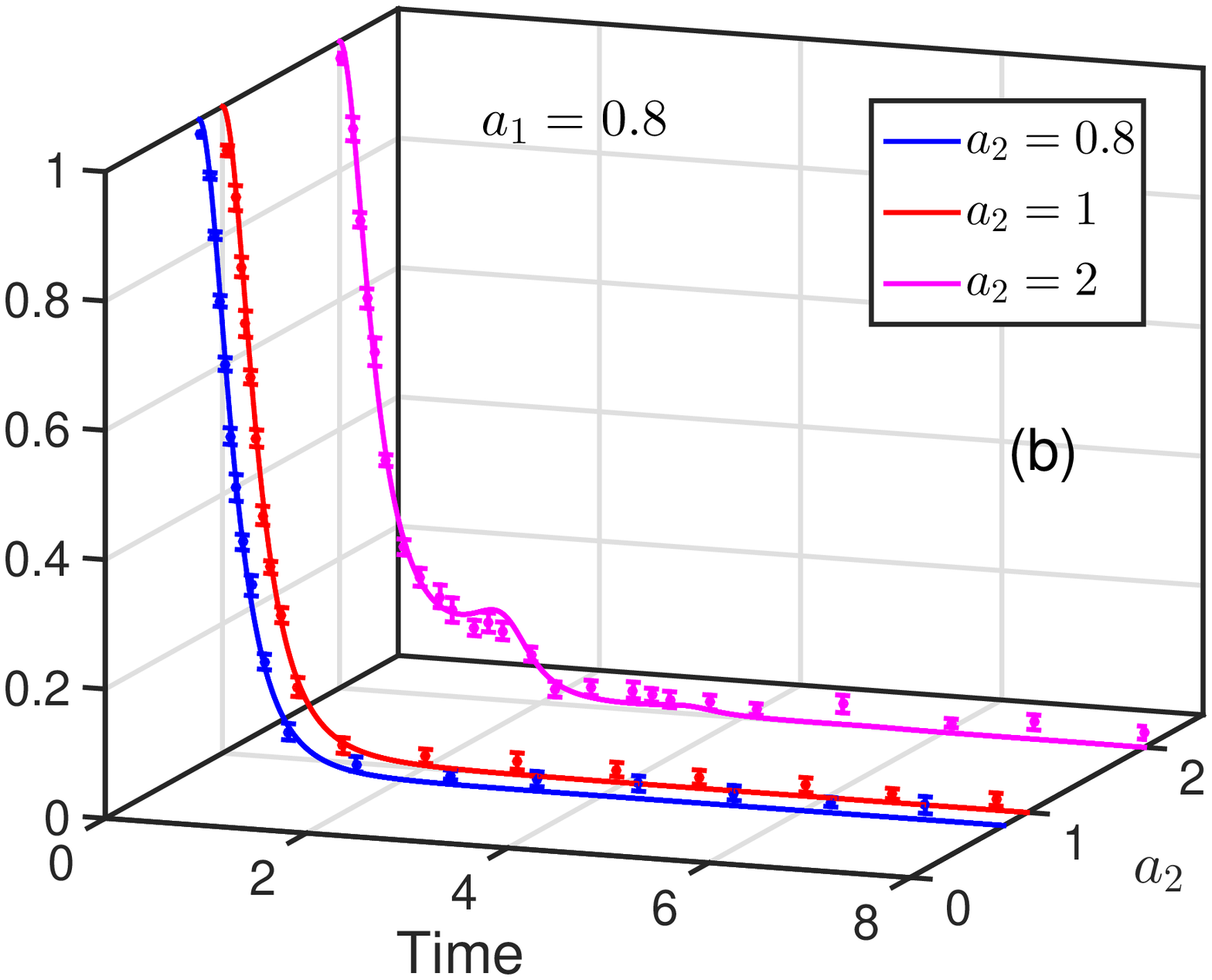}\\
    \end{flushleft}
    \vspace{1pt}
    \begin{flushleft}\hspace*{-0.4pt}
		\subfigure{}
		\includegraphics[width=0.46\linewidth]{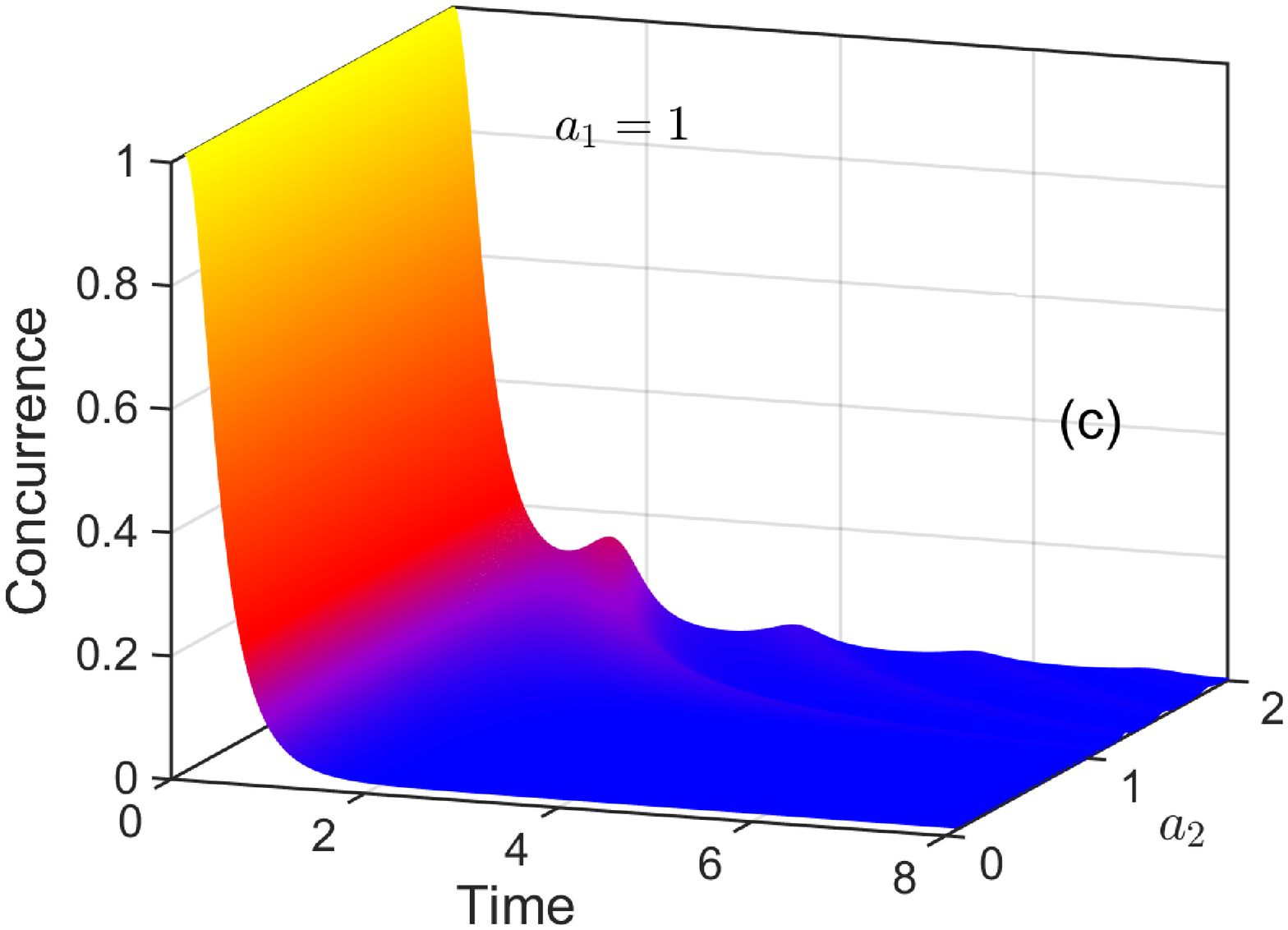}\hspace*{15pt}
		~~
		\subfigure{}
		\includegraphics[width=0.46\linewidth]{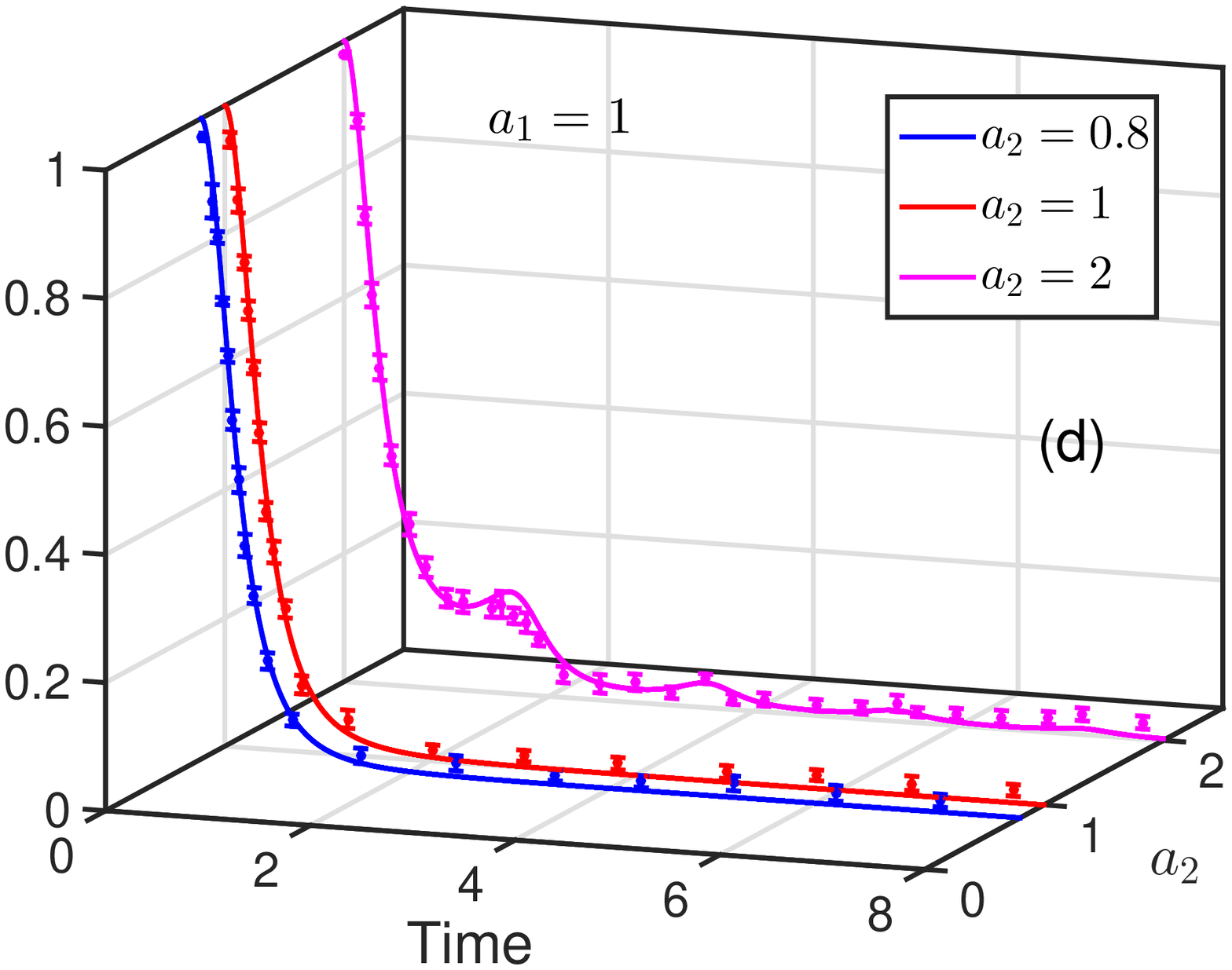}
		\caption{Time evolution of the concurrence when the first qubit evolves in the $\mathcal{APT}$-symmetric broken regime $(a_{1}=0.8)$ and at the exceptional point $(a_{1}=1.0)$, respectively. (a) and (c): theoretical simulations of the entanglement evolution with different values of $a_{2}$, when $a_{1}=0.8$ and $a_{1}=1.0$, respectively. (b) and (d): experimental and theoretical results for the entanglement evolution when $a_{2}=0.8$ (blue curve), $a_{2}=1$ (red curve), and $a_{2}=2$ (magenta curve) in the cases of $a_{1}=0.8$ and $a_{1}=1.0$, respectively. All curves show the theoretical results while the dots are experimental data.}\label{fig4}
\end{flushleft}
	\end{figure*}
\indent Figures \ref{fig4}(a) and \ref{fig4}(b) present the dynamical evolution of entanglement when qubit $\rm 1$ (the first qubit) evolves in the $\mathcal{APT}$-symmetric broken regime ($a_1=0.8$). Figure \ref{fig4}(a) is the theoretical simulation which shows a slight recovery of entanglement when $a_{2}>1$. Figure \ref{fig4}(b) shows the theoretical and experimental results when: (i) $a_{2}=0.8$ (blue curve), (ii) $a_{2}=1$ (red curve) and (iii) $a_{2}=2$ (magenta curve). One can see from Figure \ref{fig4}(b) that when $a_{2}\leq 1$, the entanglement rapidly decays to zero, while a slight recovery appears when $a_{2}>1$. The experimental results agree well with the theoretical simulation results. As $a_{2}$ increases, we have observed the phenomenon of delayed vanishing of entanglement.\\
\indent Figures~\ref{fig4}(c) and \ref{fig4}(d) present the dynamical evolution of entanglement when $a_{1}=1$, i.e., the first qubit being at the exceptional point. We investigate the entanglement evolution for different values of $a_{2}$. Figure \ref{fig4}(c) shows the theoretical simulation exhibiting multiple recoveries of entanglement when $a_{2}>1$. Figure \ref{fig4}(d) shows the theoretical and experimental results when: (i) $a_{2}=0.8$ (blue curve), (ii) $a_{2}=1$ (red curve), and (iii) $a_{2}=2$ (magenta curve). One can see from Figure \ref{fig4}(d) that when $a_{2}\leq 1$, entanglement rapidly decays to zero, which is similar to Fig. \ref{fig4}(b). While when $a_{2}>1$, a slight recovery of entanglement appears. Here, we note that with decreasing $a_{1}$ ($a_{2}$), the ratio of the Hermitian component $a_1 \sigma_{1,z}$ ($a_2\sigma_{2,z}$) to the non-Hermitian  $ i\sigma_{1,x}$ $(i\sigma_{2,x})$ component in the Hamiltonian $\hat{H}_{1,\mathcal{APT}}$ ($\hat{H}_{2,\mathcal{APT}}$) decreases. In the limiting case when $a_1 \simeq 0$ ($a_2 \simeq 0$), the system reverts to a non-Hermitian system where a strong decay of entanglement occurs. Thus, provided $a_{1}$ or $a_{2}$ is smaller than 1, i.e., one qubit evolves in the $\mathcal{APT}$-symmetric broken regime, the entanglement will eventually decay to zero, as shown in Fig. \ref{fig4}. Furthermore, the exceptional point ($a_{1}=a_{2}=1$) determines the minimal restorable entanglement. As depicted in Figs. \ref{fig2} and \ref{fig3}, when both $a_{1}$ and $a_{2}$ are larger than 1, the concurrence can always recover to a nonzero value.

\section{Discussion and conclusion}

In our experiment, the time evolution of the quantum state is realized by enforcing the nonunitary evolution operator on the quantum state at any given time, and the nonunitary operator is implemented by decomposing it into a product of unitary matrices and a loss-dependent operator. The experimental implementation of the time evolution of the quantum state is thus achieved by using optical elements. This approach can be applied to realize any nonunitary operator in $\mathcal{APT}$-symmetric systems.

The Bell state in the experiment is generated through a degenerate spontaneous parametric down-conversion, by pumping two type-II phase-matched nonlinear $\beta$-barium-borate (BBO) crystals with a 404 nm pump laser, where each BBO crystal is 0.4 mm thick and the optical axes are perpendicular to each other. The power of the pump laser is 130 mW. The quantum state is calibrated by two set of wave-plates (a half wave-plate sandwiched between two quarter wave-plates). The bandwidth of the IFs is 10 nm. This yields a maximum count of 10,000 per 1.5 s. In the measurement part, the quantum state is measured by performing standard state tomography, i.e., by projecting the state onto 16 bases \{$|HH\rangle$, $|HV\rangle$, $|VV\rangle$, $|VH\rangle$, $|RH\rangle$, $|RV\rangle$, $|DV\rangle$, $|DH\rangle$, $|DR\rangle$, $|DD\rangle$, $|RD\rangle$, $|HD\rangle$, $|VD\rangle$, $|VL\rangle$, $HL\rangle$, $|RL\rangle$\}, where $|D\rangle = \left(|H\rangle+|V\rangle\right)/\sqrt{2}$, $|R\rangle = \left(|H\rangle - i |V\rangle\right)/\sqrt{2}$, and $|L\rangle = \left(|H\rangle+i|V\rangle\right)/\sqrt{2}$. Then the density matrix is determined by carrying out a maximum-likelihood estimation algorithm. To project the qubit onto $|H\rangle$, $|V\rangle$, $|D\rangle$, $|R\rangle$, and $|L\rangle$ states, the angles of the QWP and HWP are set at $\left(0^{\circ},~0^{\circ}\right)$, $\left(0^{\circ},~45^{\circ}\right)$, $\left(45^{\circ},~22.5^{\circ}\right)$, $\left(0^{\circ},~22.5^{\circ}\right)$, and $\left(45^{\circ},~0^{\circ}\right)$, respectively.

The imperfections of the experiment are mainly caused by the instability of interference at the BDs and the inaccuracy of  angles of the wave plates. The results are also influenced by decoherence of the quantum state, incorrect relative phases inside the interferometers formed by BDs, and non-identical coupling efficiency of the two polarizations at the single-mode fibers, among other imperfections.

We have investigated the dynamics of entanglement between two qubits in an $\mathcal{APT}$-symmetric system. Our theoretical simulations demonstrate periodic oscillations, nonperiodic oscillations, rapid decay and delayed vanishing, and sudden vanishing and revival of entanglement. When both qubits evolve identically in the $\mathcal{APT}$-symmetric unbroken regime, the entanglement oscillates periodically and the peak can reach unity, which means that the entanglement is well protected in this case. When both qubits evolve differently in the $\mathcal{APT}$-symmetric unbroken regime, there exist nonperiodic oscillations of entanglement. Remarkably, when the two qubits evolve near the exceptional point in the $\mathcal{APT}$-symmetric unbroken regime, there exist the sudden vanishing and revival of entanglement.  In addition, we further theoretically study the dynamics of entanglement when only one qubit evolves under an $\mathcal{APT}$-symmetric Hamiltonian. We find that in this case, periodic oscillations of the concurrence still exist  (see Appendix D); however, the nonperiodic oscillations of concurrence, previously shown in Figs. \ref{fig3} (a), do not exist (see Appendix D).  Thus, the nonperiodic oscillations of the concurrence is a unique phenomenon in the case when both qubits evolve under under $\mathcal{APT}$-symmetric Hamiltonians. In this work, we have also performed an experiment with a linear optical setup. The experimental results agree well with our theoretical simulation results when two qubits evolve under $\mathcal{APT}$-symmetric Hamiltonians.  This work thus demonstrate the entanglement evolution in an $\mathcal{APT}$-symmetric system. The phenomena discovered in this work provide insight into the study of quantum open systems. This work opens a new door for future studies on the dynamics of quantum entanglement in multiqubit $\mathcal{APT}$-symmetric systems and other non-Hermitian quantum systems.

\section*{ACKNOWLEDGMENTS}
This work was partly supported by the Key-Area Research and Development Program of GuangDong province (2018B030326001), the National Natural Science Foundation of China (NSFC) (11074062, 11374083, 11774076, 11804228, U21A20436), the Jiangxi Natural Science Foundation (20192ACBL20051, 20212BAB211018), and the Project of Jiangxi Province Higher Educational Science and Technology Program (Grant Nos. GJJ190891, GJJ211735). F.N. is supported in part by: Nippon Telegraph and Telephone Corporation (NTT) Research, the Japan Science and Technology Agency (JST) [via the Quantum Leap Flagship Program (Q-LEAP), and the Moonshot R\&D Grant Number JPMJMS2061], the Japan Society for the Promotion of Science (JSPS) [via the Grants-in-Aid for Scientific Research (KAKENHI) Grant No. JP20H00134], the Army Research Office (ARO) (Grant No.~W911NF-18-1-0358), the Asian Office of Aerospace Research and Development (AOARD) (via Grant No. FA2386-20-1-4069), and the Foundational Questions Institute Fund (FQXi) via Grant No. FQXi-IAF19-06.


\onecolumngrid
 \renewcommand{\section}{\textbf{A.~}}
\renewcommand{\theequation}{A\arabic{equation}}
\setcounter{equation}{0}
\setcounter{subsection}{0}

\subsection*{{\textbf{\large{Appendix A: The difference of entanglement dynamics in $\mathcal{APT}$- and $\mathcal{PT}$-symmetric systems}}}}

In the main text, we point out that the dynamics in $\mathcal{APT}$-symmetric systems is different from that in  $\mathcal{PT}$-symmetric systems, even though there is a one-to-one correspondence between the Hamiltonian $\hat{H}_{\mathcal{PT}}$ in $\mathcal{PT}$-symmetric  systems and the Hamiltonian $\hat{H}_{\mathcal{APT}}$  in $\mathcal{APT}$-symmetric systems, i.e.,  $\hat{H}_{\mathcal{APT}} = i\hat{H}_{\mathcal{PT}}$, where $\hat{H}_{\mathcal{PT}}=\hat{\sigma}_{x}-\mathrm{i} a \hat{\sigma}_{z}$, $\hat{H}_{\mathcal{APT}}=\mathrm{i} \hat{\sigma}_{x}+a \hat{\sigma}_{z}$.
\begin{figure*}[!htbp]
\vspace{13pt}
\begin{flushleft}\hspace*{45pt}
 \includegraphics[scale=0.5]{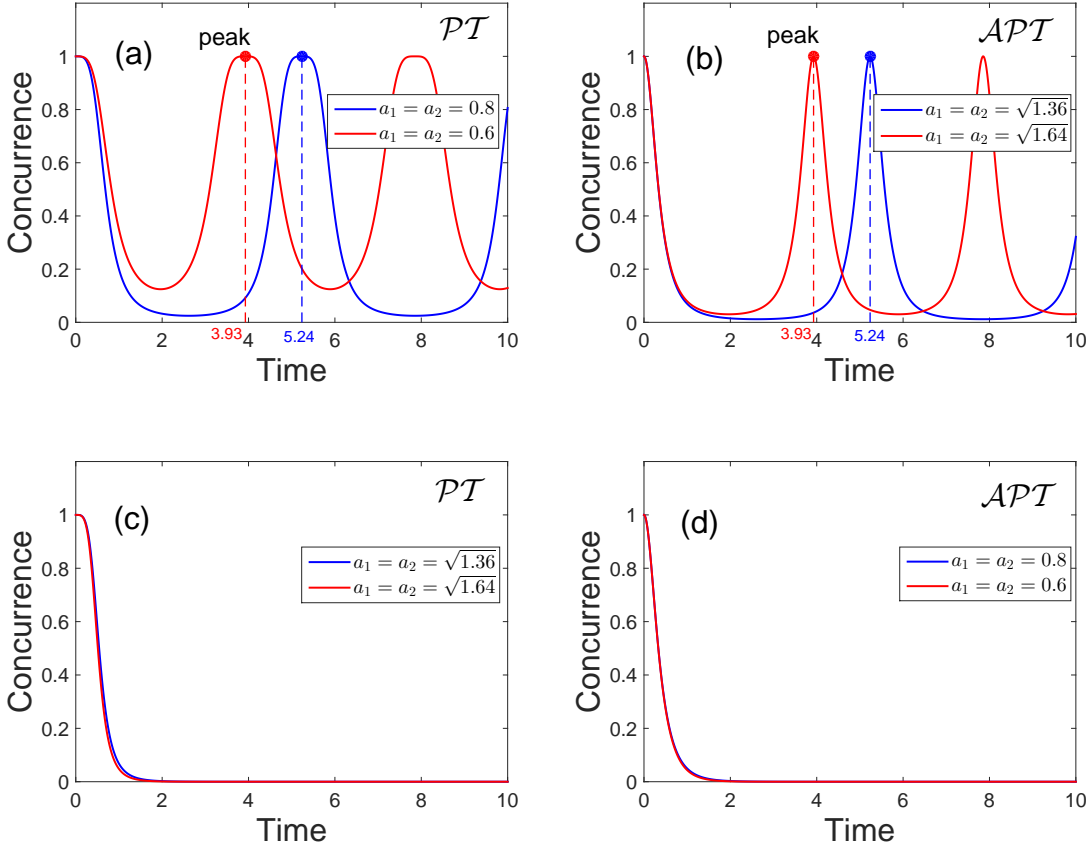}
\caption{{The concurrence dynamics for the two qubits initially in the Bell state $\frac{1}{\sqrt{2}}\left(|01\rangle +|10\rangle\right)$. (a) and (c) : The evolution of concurrence in the $\mathcal{PT}$-symmetric system. (b) and (d): The evolution of concurrence in the $\mathcal{APT}$-symmetric system. Note that the values of the parameters $a_1$ and $a_2$ are chosen here such that the oscillation periods of concurrence are the same for both $\mathcal{PT}$- and $\mathcal{APT}$-symmetric evolutions. } }
\label{afig4}
\end{flushleft}
\end{figure*}

To intuitively show the difference of entanglement dynamics in $\mathcal{APT}$- and $\mathcal{PT}$-symmetric systems, we plot the time evolution of entanglement under different parameters for $\mathcal{PT}$-symmetric [Figs.~\ref{afig4} (a, c)] and $\mathcal{APT}$-symmetric [Figs. \ref{afig4} (b, d)] systems. Both Figs. \ref{afig4} (a, b) are for the symmetry-unbroken cases, while Figs.~\ref{afig4} (c, d) are for the symmetry-broken cases. By comparing Fig.~\ref{afig4}(a) with Fig.~\ref{afig4}(b) (comparing Fig.~\ref{afig4}(c) with \ref{afig4}(d)), one can easily find that for the case of  $a_1 = a_2$, the dynamics of entanglement in $\mathcal{APT}$-symmetric systems is quite different from that in $\mathcal{PT}$-symmetric systems. For symmetry-unbroken cases, the periods for the $\mathcal{APT}$- and $\mathcal{PT}$-symmetric systems are $T_{c,\mathcal{APT}}=\frac{\pi}{\sqrt{a^2-1}}$ and $T_{c,\mathcal{PT}}=\frac{\pi}{\sqrt{1-a^2}}$, respectively. As shown in Fig.~\ref{afig4}(a) and Fig. \ref{afig4}(b), one can see that the concurrence oscillates periodically and the periods are the same for both $\mathcal{APT}$- and $\mathcal{PT}$-symmetric systems. However, in the vicinity of the peak values, the concurrence changes more dramatically under $\mathcal{APT}$ evolution [see Fig. \ref{afig4}(b)] than that under $\mathcal{PT}$ evolution [see Fig. \ref{afig4}(a)]. On the other hand, for symmetry-broken cases,  Figs.~\ref{afig4}(c) and (d) show that the entanglement dynamics in $\mathcal{APT}$- and $\mathcal{PT}$-symmetric systems are similar. Nevertheless, the entanglement drops more quickly at the beginning in $\mathcal{APT}$-symmetric systems.

\renewcommand{\theequation}{B\arabic{equation}}
\setcounter{equation}{0}
\setcounter{subsection}{0}

\subsection*{\textbf{ \large{Appendix B: Decomposition of nonunitary operators and their experimental implementations}}}

 The dynamic evolution of qubit $j$ ($j=1,~2$) is characterized by the nonunitary operator $U_{j,\mathcal{APT}}=\exp(-i \hat{H}_{j,\mathcal{APT}})$, with the non-Hermitian Hamiltonian $\hat{H}_{j,\mathcal{APT}} = i \hat{\sigma}_{j,x} + a_j \hat{\sigma}_{j, z}$. Since $H_{1, \mathcal{APT}}$ and $H_{2, \mathcal{APT}}$ take the same form, both $U_{1,\mathcal{APT}}$ and $U_{2,\mathcal{APT}}$ have the same form of decomposition and can be implemented by using the same combination of optical elements.


 Let us start with the nonunitary operator $U_{1,\mathcal{APT}}$. This operator $U_{1,\mathcal{APT}}$ can be expressed as:
 \begin{eqnarray}
   \boldsymbol{U}_{1,\mathcal{APT}}(t) & = & \exp(-i\hat{H}_{1,\mathcal{APT}}t)\nonumber\\
                           & = & \exp\left[-i(i \sigma_{1,x}+a \sigma_{1,z})t\right]\nonumber\\
     & = &\exp\left[ \left(\begin{array}{cc}
                    -ia_1 & 1\\
                    1    & ia_1
                  \end{array}
             \right)t\right]\nonumber\\
     & = &\left(\begin{array}{cc}
                    A-iB & C\\
                    C    & A+iB
                  \end{array}
             \right).\label{Uapt}
 \end{eqnarray}
Here $A$, $B$ and $C$ are given by

(i) for $a_1>1$,
\begin{equation}
 A=\cos\left(\omega_{1,1} t\right),~~B=\frac{a_1}{\omega_{1,1}}\sin\left(\omega_{1,1} t\right),~~C=\frac{1}{\omega_{1,1}}\sin\left(\omega_{1,1} t\right),\label{Eq:APTABC}
\end{equation}
where $\omega_{1,1}=\sqrt{a_1^2-1} >0$.

(ii) for $0<a_1<1$,
\begin{eqnarray}
A=\cosh\left(\omega_{1,2} t\right),~~B=\frac{a_1}{\omega_{1,2}}\sinh\left(\omega_{1,2} t\right),\nonumber\\
C=\frac{1}{\omega_{1,2}}\sinh\left(\omega_{1,2} t\right),\label{Eq:APTABC2}
\end{eqnarray}
where $\omega_{1,2}=\sqrt{1-a_1^2} >0$.

(iii) for $a_1=1$,
\begin{equation}
A=1,~~~ B=t, ~~~C=t. \label{Eq:APTABC3}
\end{equation}
  The matrix of Eq.~(\ref{Uapt}) can be rewritten as
\begin{equation}
   \boldsymbol{U}_{1, \mathcal{APT}}(t) = \frac{1}{2}\left(
    \begin{array}{cc}
     (\lambda_{1,1}+\lambda_{1,2}) e^{-2(\theta_{1,1}+\theta_{1,2})i} &  (\lambda_{1,1}-\lambda_{1,2}) e^{2(\theta_{1,1}-\theta_{1,2})i} \\
       (\lambda_{1,1}-\lambda_{1,2}) e^{-2(\theta_{1,1}-\theta_{1,2})i} &  (\lambda_{1,1}+\lambda_{1,2}) e^{2(\theta_{1,1}+\theta_{1,2})i} \\
    \end{array}
  \right),\label{UAPTsp}
 \end{equation}
with
 \begin{eqnarray}
 \lambda_{1,1} & = &\sqrt{A^2+B^2}-C,\label{lambda1}\\
 \lambda_{1,2} & = &\sqrt{A^2+B^2}+ C,\label{lambda2}\\
  \theta_{1,1} & = & \frac{1}{4}\left[\arg(A+iB)+k\pi \right],\label{theta1}\\
  \theta_{1,2} & = & \frac{1}{4}\left[\arg(A+iB)- k\pi\right],\label{theta2}
 \end{eqnarray}
 where $k$ is an  integer and $\arg(A+iB)$ denotes the argument principal value of ($A+iB$). The matrix of Eq.~(\ref{UAPTsp}) can be decomposed as follows:
 \begin{eqnarray}
   \boldsymbol{U}_{1,\mathcal{APT}}(t) & = & \frac{1}{2}\left(
    \begin{array}{cc}
     -\lambda_{1,2} e^{-2\theta_{1,2} i} &  \lambda_{1,1} e^{-2\theta_{1,2} i} \\
       \lambda_{1,2} e^{2\theta_{1,2} i} &  \lambda_{1,1} e^{2\theta_{1,2} i} \\
    \end{array}
  \right)\left(
    \begin{array}{cc}
     - e^{-2\theta_1 i} &   e^{2\theta_1 i} \\
        e^{-2\theta_1 i} &   e^{2\theta_1 i} \\
    \end{array}
  \right)\nonumber\\
  & = &\frac{1}{2}\left(
    \begin{array}{cc}
      e^{-2\theta_{1,2} i} &   -e^{-2\theta_{1,2} i} \\
      e^{2\theta_{1,2} i} &  e^{2\theta_{1,2} i} \\
    \end{array}
  \right)\left(
    \begin{array}{cc}
     0 &  \lambda_{1,1} \\
      \lambda_{1,2} & 0 \\
    \end{array}
  \right)
  \left(
    \begin{array}{cc}
     - e^{-2\theta_{1,1} i} &   e^{2\theta_{1,1} i} \\
        e^{-2\theta_{1,1} i} &   e^{2\theta_{1,1} i} \\
    \end{array}
  \right)\nonumber\\
  & = &\left(
    \begin{array}{cc}
      -e^{-2\theta_{1,2} i} &  0 \\
      0 &  e^{2\theta_{1,2} i} \\
    \end{array}
  \right)\frac{1}{\sqrt{2}}\left(
    \begin{array}{cc}
     -1 &  1 \\
      1 & 1 \\
    \end{array}
  \right)\left(
    \begin{array}{cc}
     0 &  \lambda_{1,1} \\
      \lambda_{1,2} & 0 \\
    \end{array}
  \right)\frac{1}{\sqrt{2}}\left(
    \begin{array}{cc}
     1 &  1 \\
      -1 & 1 \\
    \end{array}
  \right)
  \left(
    \begin{array}{cc}
     -e^{-2\theta_{1,1} i} &   0 \\
        0 &   e^{2\theta_{1,1} i} \\
    \end{array}
  \right).\nonumber\\
  \label{Uaptsp1}
 \end{eqnarray}
 A half-wave plate (HWP) and a quarter-wave plate (QWP) perform rotation operations, which are described by the following operators:
\begin{eqnarray}
    R_{\rm QWP}\left(\alpha\right) & = & \left(\begin{array}{cc}
                                      \cos^2 \alpha + i\sin^2 \alpha & \sin \alpha \cdot \cos \alpha \left(1-i\right) \\
                                      \sin \alpha \cdot \cos \alpha \left(1-i\right) & \sin^2 \alpha + i\cos^2 \alpha\\
                                    \end{array}
                                  \right),\label{RQWP}\\
    R_{\rm HWP}\left(\beta\right) & = & \left(\begin{array}{cc}
                                      \cos 2 \beta & \sin 2 \beta\\
                                      \sin 2 \beta & -\cos 2 \beta\\
                                    \end{array}
                                  \right),\label{RHWP}
   \end{eqnarray}
where $\alpha$ and $\beta$ are tunable setting angles. Based on Eq.~(\ref{RQWP}) and Eq.~(\ref{RHWP}), we have:
\begin{eqnarray}
 R_{\rm QWP}\left(45^{\circ}\right)R_{\rm HWP}\left(\theta_{1,j}\right) R_{\rm QWP}\left(45^{\circ}\right) & = & \left(
    \begin{array}{cc}
     -e^{-2\theta_{1,j} i} &   0 \\
        0 &   e^{2\theta_{1,j} i} \\
    \end{array}
  \right),~~(j=1,2),\label{HWPz}\\
  R_{\rm HWP}\left(0^{\circ}\right)R_{\rm HWP}\left(22.5^{\circ}\right) & = & \frac{1}{\sqrt{2}}\left(\begin{array}{cc}
     1 &   1 \\
        -1 &   1 \\
    \end{array}
  \right),\label{HWP0}\\
  R_{\rm HWP}\left(67.5^{\circ}\right) & = & \frac{1}{\sqrt{2}}\left(\begin{array}{cc}
     -1 &   1 \\
        1 &   1 \\
    \end{array}
  \right).\label{HWP67}
\end{eqnarray}
 After inserting Eqs.~(\ref{HWPz})-(\ref{HWP67}) into Eq.~(\ref{Uaptsp1}), we obtain:
\begin{equation}
 U_{1,\mathcal{APT}}=U_{1,2}(\theta_{1,2}) \mathcal{M} U_{1,1}(\theta_{1,1}),\label{uasp}
\end{equation}
with
 \begin{eqnarray}
 U_{1,1}(\theta_{1,1}) & = & R_{\rm HWP}(0^{\circ})R_{\rm HWP}(22.5^{\circ})R_{\rm QWP}(45^{\circ}) R_{\rm HWP}(\theta_{1,1}) R_{\rm QWP}(45^{\circ}),\label{eq:u1}\\
 U_{1,2}(\theta_{1,2}) & = & R_{\rm QWP}(45^{\circ}) R_{\rm HWP}(\theta_{1,2})R_{\rm QWP}(45^{\circ}) R_{\rm HWP}(67.5^{\circ}),\label{eq:u2}\\
 \mathcal{M} & = & \left(\begin{array}{cc} 0 &  \lambda_{1,1} \\ \lambda_{1,2} & 0  \end{array}\right).\label{eq:M}
 \end{eqnarray}

 The matrix $\mathcal{M}$ can be expressed as:
\begin{equation}
\mathcal{M}=c\left(
   \begin{array}{cc}
     0 & \sin 2\xi_{1,1} \\
     \sin 2\xi_{1,2} & 0 \\
   \end{array}
 \right),
\end{equation}
where $c=\frac{\lambda_{1,1}}{\sin{2\xi_{1,1}}}=\frac{\lambda_{1,2}}{\sin{2\xi_{1,2}}}$ is a trivial constant. For simplicity, we define:
\begin{equation}
L_1\left(\xi_{1,1}, \xi_{1,2}\right)=\left(
\begin{array}{cc}0 &  \sin 2\xi_{1,1} \\
\sin 2\xi_{1,2} & 0 \end{array}\right).\label{loss}
\end{equation}
Thus, we have $\mathcal{M}=cL_1$. Note that the functions of both operators $L_1$ and $cL_1$ are identical. This is because the states $L_1 | \psi\rangle $ and $cL_1 |\psi\rangle$, obtained by enforcing the two operators $L_1$ and $cL_1$ on an arbitrary  state $|\psi\rangle$, are the same according to the principles of quantum mechanics. Therefore, we can replace $\mathcal{M}$ in Eq.~(\ref{uasp}) by the operator $L_1$. In this sense, we have from Eq.~(\ref{uasp}):\\
\begin{equation}
 U_{1,\mathcal{APT}}=U_{1,2}(\theta_{1,2}) L_1\left(\xi_{1,1},~\xi_{1,2}\right) U_{1,1}(\theta_{1,1}).\label{uasp1}
\end{equation}
which is exactly the same as the decomposition of the nonunitary operator $U_{1,\mathcal{APT}}$, described by Eq.~(\ref{Eq:UaptE}) in the main text (with $j=1$).

\begin{figure}[!htbp]
\begin{flushleft}\hspace*{0pt}
 \includegraphics[width=1\linewidth]{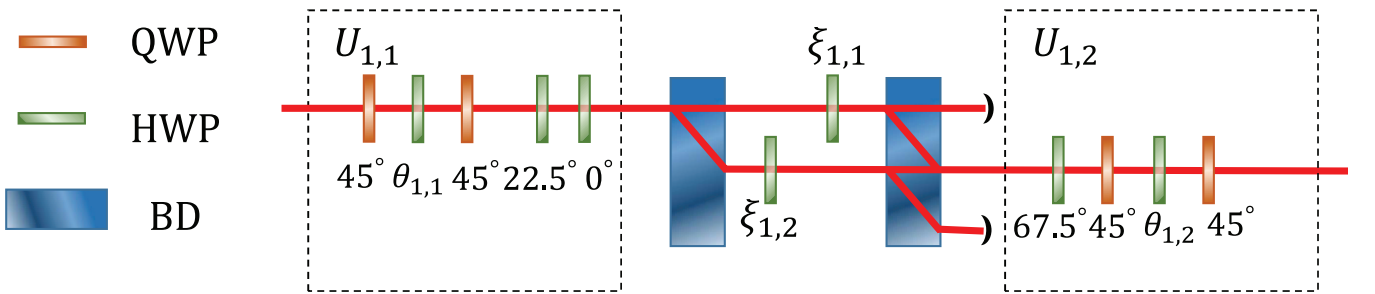}
\caption{Experimental setup to realize the nonunitary operator $U_{1,\mathcal{APT}}(t)$. Here, the wave plates in the two dashed rectangles are used to realize the operators $U_{1,1}$ and $U_{1,2}$ (see Eqs.~(\ref{eq:u1}) and (\ref{eq:u2})), while the two BDs with two HWPs inside are used to realize the loss operator $L_1$. Note that this setup is the same as the upper layer optical elements in the center area of Fig.~\ref{fig1} in the main text.}
\label{afig1}
\end{flushleft}
\end{figure}

\begin{figure}[!htbp]
 \includegraphics[scale=0.9]{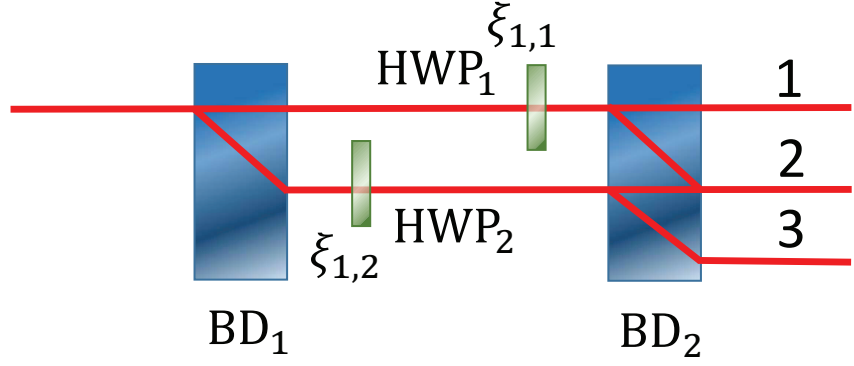}
\caption{Optical implementation for the loss operator $L_1$, where $\xi_{1,1}$ and $\xi_{1,2}$ are the two tunable setting angles for the half-wave plates $\rm HWP_1$ and $\rm HWP_2$, respectively.}
\label{afig2}
\end{figure}

In the following, we show the optical implementation of $U_{1,\mathcal{APT}}$. According to Eqs.~(\ref{eq:u1}) and (\ref{eq:u2}), the operators $U_{1,2}(\theta_{1,2})$ and $U_{1,1}(\theta_{1,1})$ can be straightforwardly realized by using HWPs and QWPs (Fig.~\ref{afig1}). We realize the loss operator $L_1$ by using a combination of two beam displacers (BDs) and two HWPs (Fig.~\ref{afig2}).

In Fig. \ref{afig2}, the optical axes of the BDs are cut so that the vertically polarized photons are transmitted directly, while the horizontally polarized photons are displaced into the lower path. In addition, $\rm HWP_1$ and $\rm HWP_2$ with setting angles $\xi_{1,2}$ and $\xi_{1,2}$ are, respectively, inserted into the upper and lower paths between the two BDs. The rotation operations on the photon polarization states, performed by the $\rm HWP_1$ and $\rm HWP_2$, are given as follows:

\begin{equation}
R_{\rm HWP}\left(\xi_{1,1}\right)=\left(
                             \begin{array}{cc}
                               \cos 2\xi_{1,1} & \sin 2\xi_{1,1} \\
                               \sin 2\xi_{1,1} & -\cos 2\xi_{1,1} \\
                             \end{array}
                           \right),~~~~
 R_{\rm HWP}\left(\xi_{1,2}\right)=\left(
                             \begin{array}{cc}
                               \cos 2\xi_{1,2} & \sin 2\xi_{1,2} \\
                               \sin 2\xi_{1,2} & -\cos 2\xi_{1,2} \\
                             \end{array}
                           \right).
\end{equation}
In this case, when photons with different polarizations pass through the setup in Fig.~\ref{afig2}, they are transformed as follows:
\begin{eqnarray}
|H\rangle~\xrightarrow{\rm BD_1}~|H\rangle_{\rm lower}~\xrightarrow{R_{\rm HWP}(\xi_{1,2})}~R_{\rm HWP}(\xi_{1,2})|H\rangle ~\xrightarrow{\rm BD_2}~\cos 2\xi_{1,2}|H\rangle_3 +\sin 2\xi_{1,2} |V\rangle_2,\label{BDH}\\
|V\rangle ~ \xrightarrow{\rm BD_1} ~ |H\rangle_{\rm upper} ~ \xrightarrow{R_{\rm HWP}(\xi_{1,1})}~ R_{\rm HWP}(\xi_{1,1})|V\rangle ~ \xrightarrow{\rm BD_2} ~ \sin 2\xi_{1,1}|H\rangle_2 -\cos 2\xi_{1,1} |V\rangle_1,\label{BDV}
\end{eqnarray}
where the subscript ``lower'' represents the lower path between the two BDs, while the subscripts ``1'', ``2'', and ``3'' represent the three paths after the second BD. Only horizontally polarized photons in the upper path and vertically polarized photons in the lower path are transmitted through the second BD and then combined into path $\rm 2$, while other photons transmitted into path $\rm 1$ or $\rm 3$ are blocked, i.e., they are lost from the system. In this sense, according to Eqs. (\ref{BDH}) and (\ref{BDV}), one can easily find that when the input photon is  initially in the state $\rho_{\rm in}=\sum p_j |\psi_j\rangle \langle \psi_j|$, where $|\psi_j\rangle = \alpha_j|H\rangle + \beta_j e^{i\varphi_j}|V\rangle$, then the output photon appearing in the path $2$ would be in the state $\rho_{\rm out}=\sum p_j |\psi^{\prime}_j\rangle \langle \psi^{\prime}_j|$, where $|\psi^{\prime}_j\rangle = \alpha_j \sin 2 \xi_{1,2} |V\rangle + \beta_j e^{i\varphi_j} \sin 2\xi_{1,1}|H\rangle$. It is obvious that this state transformation can be written as $|\psi_j^{\prime}\rangle=L_1|\psi_j\rangle$, i.e., $\rho_{\rm out}=L_1 \rho_{\rm in} L_1^{\dagger}$, which implies that the setup in Fig.~\ref{afig2} realizes the loss-dependent operator $L_1$ given in Eq.~(\ref{loss}).

In the above, we have proved the decomposition of the nonunitary  operator $U_{1,\mathcal{APT}}$ in Eq.~(\ref{Eq:UaptE}) in the main text. We have also discussed the experimental realization of this nonunitary operator. Since the Hamiltonian $H_{2,\mathcal{APT}}$ takes the same form as the Hamiltonian $H_{1,\mathcal{APT}}$, the proof of the decomposition of the nonunitary operator $U_{2,\mathcal{APT}}$ for qubit $\rm 2$ ($j=2$) is the same as above. In addition, since  $U_{2,\mathcal{APT}}$ has a similar decomposition form as $U_{1,\mathcal{APT}}$, it is straightforward to see that the nonunitary operator $U_{2,\mathcal{APT}}$ can be realized with the lower layer optical elements in the gray part of Fig.~\ref{fig1} in the main text.\\

\renewcommand{\theequation}{C\arabic{equation}}
\setcounter{equation}{0}
\setcounter{subsection}{0}
\subsection*{{\textbf{\large{Appendix C: Period of concurrence under $\mathcal{APT}$-symmetric evolution}}}}
{Let us consider the case when $a_1,~a_2 >1$ (i.e., the $\mathcal{APT}$-symmetric unbroken regime). The two qubits are initially in a Bell state $|\psi_0\rangle=\frac{1}{\sqrt{2}}\left(|01\rangle+|10\rangle\right)$. After $\mathcal{APT}$-symmetric evolution, the output state of the two qubits is:}
{\begin{equation}
|\psi \left(t\right) \rangle= \left(U_{1,\mathcal{APT}}\otimes U_{2,\mathcal{APT}}\right) |\psi_0\rangle,
\end{equation}}
{where $U_{j,\mathcal{APT}}=e^{-i\hat{H}_{j,\mathcal{APT}}t}$ ($j=1, 2$).}

{When $a_1=a_2=a>1$, the output state can be written by:}
{\begin{equation}
 |\psi \left(t\right) \rangle= \frac{1}{\sqrt{2}}\left(\begin{array}{c}
C_0\\
  C_1\\
  C_2\\
  C_3\\
                                                          \end{array}
                                                        \right), \label{Eq.entphi}
\end{equation}}
{in the computational basis $\{|00\rangle,~|01\rangle,~|10\rangle, |11\rangle\}$. Here, $C_0= \sin 2 \omega t /\omega -  2a\sin^2 \omega t /{\omega^2} $, $  C_1=C_2=\cos^2 \omega t + {(a^2+1)}\sin^2\omega t / {\omega^2} $, and $C_3=\sin 2 \omega t /{\omega}  +  {2a} \sin^2 \omega t/ {\omega^2}$, with $\omega=\sqrt{a^2-1}$. Note that, the state in Eq.~(\ref{Eq.entphi}) is not normalized, since the nonunitary time-evolution generated by an $\mathcal{APT}$-symmetric Hamiltonian induces loss. Nevertheless, it can be normalized as:}

{\begin{equation}
 |\psi \left(t\right) \rangle= \frac{1}{\sqrt{M}}\left(\begin{array}{c}
 C_0\\
C_1\\
 C_2 \\
 C_3\\
                                                          \end{array}
                                                        \right), \label{Eq.entphinor}
\end{equation}
where
\begin{equation}
M= \frac{2(a^2-\cos 2 \omega t)^2}{(a^2-1)^2}+\frac{8\sin^2\omega t(a^2-\cos^2\omega t)}{(a^2-1)^2}.
\end{equation}
After a simple calculation, one can find that for the state in Eq.~(\ref{Eq.entphinor}), the concurrence becomes:
\begin{eqnarray}
 C(|\psi(t)\rangle) & = & \frac{2(a^2-1)}{a^4+2a^2-4a^2\cos 2\omega t + 2\cos^2 2\omega t-1}.\label{Eq.concurrence}
\end{eqnarray}
According to Eq.~(\ref{Eq.concurrence}), one can readily see that the period of concurrence under $\mathcal{APT}$-symmetric evolution becomes:
\begin{equation}
 T_{c,APT}=\frac{2\pi}{2\omega}=\frac{\pi}{\sqrt{a^2-1}},
\end{equation}
 which shows that the period of the concurrence decreases as the parameter $a$ increases.}

\renewcommand{\theequation}{D\arabic{equation}}
\setcounter{equation}{0}
\setcounter{subsection}{0}
\subsection*{{\textbf{\large{Appendix D: The concurrence dynamics when only one qubit evolves under an $\mathcal{APT}$-symmetric Hamiltonian}}}}
\begin{figure*}[!htbp]
\vspace{10pt}
 \setlength{\abovecaptionskip}{10pt}
\begin{flushleft}\hspace*{15pt}
 \includegraphics[scale=0.5]{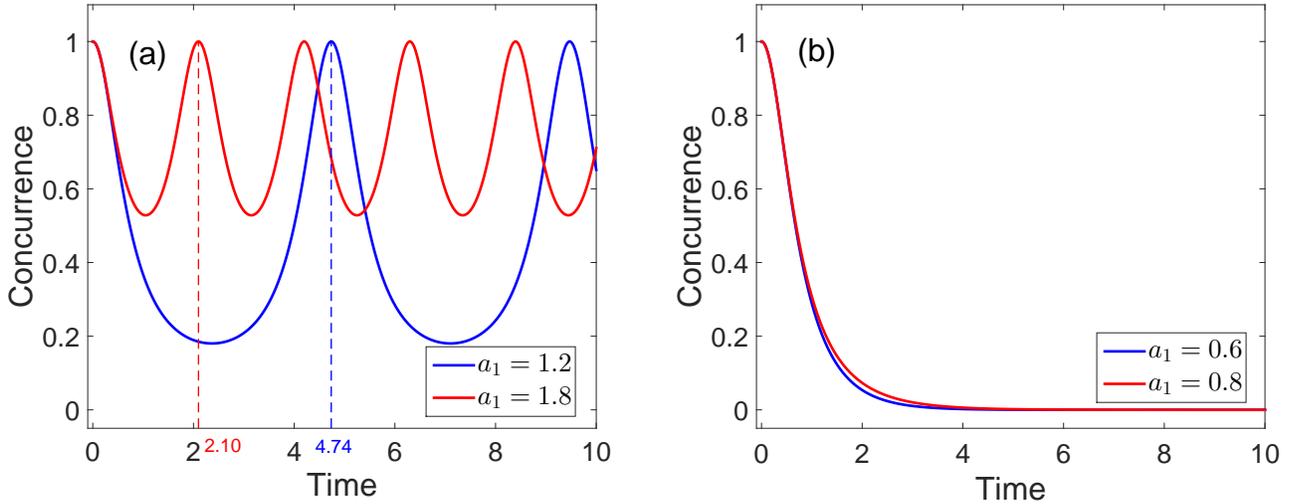}
\caption{{The concurrence dynamics when only one qubit (the first qubit) evolves under an $\mathcal{APT}$-symmetric Hamiltonian. The two qubits are initially in the Bell state $\frac{1}{\sqrt{2}}\left(|01\rangle +|10\rangle\right)$. (a) Periodic oscillation of the concurrence versus time. (b) Rapid decay of concurrence over time.}}
\label{afig5}
\end{flushleft}
\end{figure*}
In the main text, we have discussed the concurrence dynamics when both qubits evolve under $\mathcal{APT}$-symmetric Hamiltonians. Here, we  briefly discuss the dynamics of the concurrence when only one qubit (say the first qubit) evolves under an $\mathcal{APT}$-symmetric Hamiltonian. Figure \ref{afig5}(a) shows that when $a_1>1$  the concurrence oscillates periodically over time, while Fig. \ref{afig5} (b) shows that the concurrence rapidly decays over time when $a_1<1$.

Upon comparing Fig.~\ref{afig5}(a) with Fig.~\ref{fig2}(a) and comparing  Fig.~\ref{afig5}(b) with Fig.~\ref{fig2}(b), we can confirm that the entanglement dynamics of the $\mathcal{APT}$-symmetric system where only one qubit evolves under an $\mathcal{APT}$-symmetric Hamiltonian is  similar to that when both qubits evolve under the same $\mathcal{APT}$-symmetric Hamiltonian (i.e., $a_1=a_2$).  That is, when only one qubit evolves under an $\mathcal{APT}$-symmetric Hamiltonian, the concurrence exhibits either periodic oscillation or rapid decay, which only depends on the value of $a_1$. However, we should note that the minimum value of the concurrence in this case is different from the minimum value of the concurrence when both qubits evolve under the same $\mathcal{APT}$-symmetric Hamiltonian, which can be seen from the red and blue curves  in Fig. \ref{afig5}(a) and Fig. \ref{fig2}(a). Furthermore,  both Fig. \ref{afig5}(a) and Fig. \ref{afig5}(b) show that the nonperiodic oscillations of concurrence do not occur when only one qubit evolves under an $\mathcal{APT}$-symmetric Hamiltonian, which however exist in the case when both qubits evolve under different $\mathcal{APT}$-symmetric Hamiltonians, as shown in  Fig.~\ref{fig3}(a).

\twocolumngrid

\end{document}